\documentclass{article}
\usepackage[a4paper, margin=1in]{geometry}
\usepackage{graphicx} 
\usepackage{amsmath}
\usepackage{hyperref} 
\usepackage{authblk}
\usepackage{comment}
\usepackage{setspace}
\usepackage{tikz}
\usepackage{lipsum}
\usepackage{amsfonts}

\usepackage{enumitem}
\graphicspath{{Figures/}}
\usepackage{caption}
\usepackage{float}
\usepackage{parskip} 
\usepackage[skip=0pt]{caption} 
\usepackage{pdflscape}
\usepackage{graphicx}
\usepackage{authblk}

\title{Optimizing Qubit Mapping with Quasi-Orthogonal Space-Time Block Codes and Quaternion Orthogonal Designs }
\author[1,2]{Valentine Nyirahafashimana}
\author[1,3]{Nurisya Mohd Shah}
\author[4]{ Umair Abdul Halim}
\author[1,5]{Mohamed Othman}
\author[1,6]{Sharifah Kartini Said Husain}
\affil[1]{Institute for Mathematical Research (INSPEM), Universiti Putra Malaysia (UPM), 43400 UPM Serdang, Selangor, Malaysia}
\affil[2]{Kigali Independent University (ULK), Polytechnic Institute, Kigali Campus, 102 KG 14 Ave Gisozi, Rwanda}
\affil[3]{Department of Physics, Faculty of Science, Universiti Putra Malaysia, 43400 UPM Serdang, Selangor, Malaysia}
\affil[4]{Centre of Foundation Studies in Science, Universiti Putra Malaysia, 43400 UPM Serdang, Selangor, Malaysia}
\affil[5]{Department of Communication Technology and Network, Universiti Putra Malaysia, 43400 UPM Serdang,  Selangor, Malaysia}
\affil[6]{Department of Mathematics and Statistics, Faculty of Science, Universiti Putra Malaysia, 43400 UPM Serdang, Selangor, Malaysia}
\affil[*]{Corresponding author: risya@upm.edu.my; nyirahafashimanav@ulk.ac.rw}

\date{} 

\begin{document}

\maketitle

\begin{abstract}
  This study explores the qubit mapping through the  integration of Quasi-Orthogonal Space-Time Block Codes (QOSTBCs) with Quaternion Orthogonal Designs (QODs) in quantum error correction (QEC) frameworks. QOSTBCs have gained prominence for enhancing performance and reliability in quantum computing and communication systems. These codes draw on stabilizer group formalism and QODs to boost error correction, with QOSTBCs mapping logical qubits to physical ones, refines error handling in complex channels environments. Simulations results demonstrate the effectiveness of this approach by comparing the percentage improvement under various detected and corrected error conditions for four different cases, \textbf{$Z_1$} up to \textbf{$Z_4$}. The obtained simulations and implemental results show that QOSTBCs consistently achieve a higher correction improvement percentage than stabilizer Group for \textbf{$Z_1$}, \textbf{$Z_2$}, and \textbf{$Z_4$}; QOSTBCs can correct more errors than those detected, achieving over 100\% correction rates for first two cases, which indicates their enhanced resilience and redundancy in high-error environments. While for \textbf{$Z_3$}, stabilizer consistently remains above that of QOSTBCs, reflecting its slightly better performance. These outcomes indicate that QOSTBCs are reliable in making better logarithmic efficiency and error resilience, making them a valuable asset for quantum information processing and advanced wireless communication.   
\end{abstract}
\section{Introduction}

Quantum computing revolutionizes information processing and communication, with qubits as its fundamental units, relying on advanced mathematical and coding techniques for reliability and efficiency~\cite{nielsen2010quantum, bhat2022quantum}. Quasi-orthogonal codes, commonly used in code division multiple access, upgrade signal separation and system capacity by balancing orthogonality and flexibility under varying channel conditions~\cite{galli2008novel}. These codes are important in quantum communication, encoding and decoding qubits while applying group theory and its practical application like stabilizer group formalism to optimize transformations and reduce error rates. Furthermore, Quaternion Orthogonal Designs (QODs) increase fault tolerance and stability by using quaternion algebra to generate orthogonal sequences for stable error correction~\cite{seberry2007theory, mushtaq2017novel, voight2021quaternion}. By expanding a limited set of qubits into a larger, versatile qubit space, quaternion frameworks streamline mapping processes, improving error correction and encoding capacity~\cite{rau2009mapping}. The integration of group theory further optimizes qubit manipulation, enhancing quantum computing capability to perform work such as simulating complex systems, solving optimization problems, and executing computations beyond classical limits~\cite{aaronson2004improved,mermin2007quantum}. Otherwise, the stabilizer theory provides an important framework for scaling quantum systems by mapping logical qubits to larger sets of physical qubits, a key requirement for fault-tolerant quantum computation. Through stabilizer codes, such as surface codes, it enables efficient error detection and correction, ensuring flexibility and operational fidelity. This approach underpins large-scale quantum algorithms by reducing error rates and improving computational reliability, drawing on interdisciplinary insights from mathematics, computer science, and physics to drive transformative advances in computation and communication~\cite{ nielsen2010quantum,marques2022logical, gottesman1997stabilizer, head2020quantum}.

In multiple-antenna wireless communication systems, space-time block codes (STBCs) are essential for achieving full diversity and low-complexity encoding and decoding in setups with \( N \) transmit and \( M \) receive antennas~\cite{santumon2012space, de2005generalized, jafarkhani2005space}. Orthogonal STBCs (OSTBCs), when combined with additional coding schemes, form super-orthogonal codes that boost system performance. An alternatively, Quasi-OSTBCs (QOSTBCs) extend the conventional framework by offering improved diversity, scalability, modulation independence, and constructibility, especially in multiple-input multiple-output (MIMO) systems~\cite{de2005generalized,jafarkhani2005space}. These codes optimize capacity and reliability by maximizing data rates and spatial diversity while reducing interference, making them well-suited for noisy, high-error environments~\cite{jafarkhani2001quasi, tarokh1999space, seema2024large}. Additionally, QODs applies quaternion algebra to generate orthogonal signals in four-dimensional space, further intensifying reliability, diversity, and stability in signal transmission~\cite{ali2021quaternion, wu2011low}. In quantum computing, quantum error correction codes (QECCs) such as stabilizer codes and surface codes are excel in managing errors in different environments. Stabilizer codes are highly effective in low-error scenarios, while surface codes, though requiring a large number of qubits, provide strong fault tolerance for high-error conditions~\cite{terhal2015quantum, kasirajan2021quantum}. These QECCs complement QOSTBCs by effectively addressing errors in both low- and high-error regimes. QOSTBCs, with their quasi-orthogonal structure, further ensures stability and reliability in quantum communication and processing, making them particularly suited for noisy, high-error environments that require advanced error correction~\cite{cuvelier2021quantum}. 
Their adaptability makes them critical for applications like Quantum Key Distribution (QKD), fault-tolerant computing, and quantum cryptography, where error resilience and data integrity are paramount. Compared to traditional STBCs, which are effective in low-error conditions, QOSTBCs better handle complex error patterns, making them essential for scalable, high-error quantum technologies due to its orthogonality properties~\cite{badic2005space, ni2012rotated, cuvelier2023enabling, wang2024analysis, liu2004application, katsuki2023noncoherent}.

This work proposes a mathematical framework for optimizing the mapping of \(N\) logical qubits onto \(M\) physical qubits, correcting up to \(P\) errors. By combining QOSTBCs via QODs and stabilizer group formalism, it aims to develop fault rectification and reduces correction time while preserving orthogonality. It utilizes quantum computing, MIMO systems, and quaternion algebra to validate the approach through analysis, simulations, and comparative analysis, demonstrating enhanced mapping efficiency and improved error detection and correction rates across various cases. The study is organized as follows: Section~\ref{SEC II} presents a mathematical framework that optimizes the process of qubits mapping through a combined approach using QOSTBCs via QODs, and stabilizer group formalism principles in QEC. Equations (~\ref{gt}) and~(\ref{qostc}) describe the mapping from the initial state \(|\psi\rangle\) of \(N\) logical qubits to the final state \(|\phi\rangle\) on \(M\) physical qubits. Section~\ref{SEC III} discusses numerical simulations results, highlighting the percentage enhancement in error detection and correction during qubit transformation comparison. Section~\ref{SEC IV} 
presents a summary of key findings and concludes this study.

\section{Methods}\label{SEC II}

This section describes a mathematical formalism to optimize the mapping of \(N\) logical qubits to \(M\) physical qubits, enabling the correction of \(P\) errors. By integrating QOSTBCs with QODs and leveraging concepts from stabilizer group, QEC, MIMO systems, and quaternion algebra, the framework ensures efficient mapping, increased errors adjustment, and reduced correction time. The proposed method is evaluated through theoretical analysis and simulations under various error conditions, focusing on four mapping cases, represented as \( Z_i \) ( i = 1, 2, 3, and 4 ):
\begin{enumerate}
\item[\textbf{\( Z_1 \):}] Encoding 3 qubits into 8 qubits, with the ability to correct a single error.
\item[\textbf{\( Z_2 \):}] Transforming 4 qubits into 10 qubits, enabling the correction of one error.
\item[\textbf{\( Z_3 \):}] Mapping 1 qubit to 13 qubits, capable of correcting 2 errors.
\item[\textbf{\( Z_4 \):}] Transforming 1 qubit into 29 qubits, allowing the correction of up to 5 errors.
\end{enumerate}

\subsection{Qubit State Spaces, Quaternion Designs, and Complex Numbers in Quantum Processes}

In quantum computing, qubits, the fundamental units of quantum information, exist in superpositions of basis states, \( |0\rangle \) and \( |1\rangle \), unlike classical bits that can only exist in one state at a time~\cite{nielsen2010quantum}. These qubits are represented in a Hilbert space, with the initial \(N\)-qubit state space denoted as \( \mathcal{H}_N \), and the final \(M\)-qubit state space as \( \mathcal{H}_M \). The transformation between these states is governed by a unitary operator \( U \), \( U: \mathcal{H}_N \rightarrow \mathcal{H}_M \), which maps from the initial to the final qubit space and can be decomposed into orthogonal and quaternionic matrices, as:

\begin{equation}
 U = Q_k \cdot O_l \cdot Q_{k-1} \cdot O_{l-1} \cdot \ldots \cdot Q_1 \cdot O_1.
\end{equation}
Here, the Hilbert spaces \( \mathcal{H}_N \) and \( \mathcal{H}_M \) are spanned by basis states \( |0\rangle, |1\rangle, \ldots, |2^N-1\rangle \) and \( |0\rangle, |1\rangle, \ldots, |2^M-1\rangle \), respectively, with each basis state expressed as a tensor product of individual qubit states i.e., 

\begin{equation}
|\psi \rangle = |\psi_1\rangle \otimes |\psi_2\rangle \otimes \ldots \otimes |\psi_N\rangle, \quad \text{and} \quad
|\phi \rangle = |\phi_1\rangle \otimes |\phi_2\rangle \otimes \ldots \otimes |\phi_M\rangle.
\end{equation}
Error correction in quantum systems involves encoding the initial state into a larger space using a unitary transformation \( U_{\text{enc}} \), which is represented as an encoding matrix. 

\begin{equation}
|\phi\rangle = U_{\text{enc}} |\psi_i\rangle,
\end{equation}

where \( U_{\text{enc}} \) is a matrix of dimension \( M \times N \), defined as:

\begin{equation*}
U_{\text{enc}} = \begin{bmatrix}
u_{11} & u_{12} & \cdots & u_{1N} \\
u_{21} & u_{22} & \cdots & u_{2N} \\
\vdots & \vdots & \ddots & \vdots \\
u_{M1} & u_{M2} & \cdots & u_{MN} \\
\end{bmatrix}.
\end{equation*}

The encoded state \( |\phi\rangle \) integrates the original quantum state with redundancy qubits \(|\text{red}\rangle\), forming a tensor product denoted as \begin{equation}
|\phi\rangle = |\psi_i\rangle \otimes |\text{red}\rangle.
\end{equation} These redundancy qubits play a leading role in error detection and correction, as exemplified by Shor's 9-qubit code~\cite{shor1995scheme}. Further refinement error adjustment involves the use of QODs with QOSTBCs, which combine quaternionic matrices and orthogonal designs to encode quantum information in multiple qubits, with the encoding matrix \( S \) represented by elements \( s_{ij} \). These matrices are represented as:

\begin{equation}
S = \begin{bmatrix}
s_{11} & s_{12} & \cdots & s_{1N} \\
s_{21} & s_{22} & \cdots & s_{2N} \\
\vdots & \vdots & \ddots & \vdots \\
s_{M1} & s_{M2} & \cdots & s_{MN}
\end{bmatrix}.
\end{equation}

In the quaternionic Hilbert space, this matrix is expressed as:

\begin{equation}
 S = \begin{bmatrix}
s_{11} & s_{12}^* & \cdots & s_{1N}^* \\
s_{21} & s_{22}^* & \cdots & s_{2N}^* \\
\vdots & \vdots & \ddots & \vdots \\
s_{M1} & s_{M2}^* & \cdots & s_{MN}^*
\end{bmatrix},
\end{equation}
 where \( s_{ij}^* \) denotes the complex conjugate of \( s_{ij} \). This matrix, characterized by \( S \in \mathbb{C}^{M \times N} \), exhibits properties like additivity and homogeneity in the encoding process:

\begin{equation}
\begin{aligned}
&S(\alpha |\psi_1\rangle + \beta |\psi_2\rangle) = \alpha S|\psi_1\rangle + \beta S|\psi_2\rangle, \\
&S(\alpha |\psi\rangle) = \alpha S|\psi\rangle,  \quad \text{for all}  \quad \alpha, \beta \in \mathbb{C}. 
 \end{aligned}
\end{equation}

The encoded quantum state \( |\phi\rangle \) is expressed as:

\begin{equation}
|\phi\rangle = S |\psi\rangle,
\end{equation}

Otherwise, for decoding, the inverse transformation is applied using the Hermitian conjugate \( S^\dagger \), recovering the original state as:

\begin{equation}
|\psi\rangle = S^\dagger |\phi\rangle.
\end{equation}
Moreover, QOSTBCs rise error tolerance by ensuring orthogonal mappings of qubit states, optimizing quantum state transformation, and reducing communication errors. By incorporating complex numbers and quaternions, QEC effectively manages noise and errors in quantum systems. An arbitrary quantum state \( |\psi\rangle \) is represented as a linear combination of basis states with complex coefficients:  
\begin{equation}
|\psi\rangle = \sum_{i=0}^{2^N - 1} \alpha_i |i\rangle,
\end{equation}
where \( \alpha_i \) are complex amplitudes, and \( \sum_{i=0}^{2^N - 1} |\alpha_i|^2 = 1 \). The Hadamard (\(H\)) and \(X\) gates, using complex and quaternionic algebra, facilitate precise quantum state manipulation, enabling superposition creation and qubit flipping key operations for advanced quantum information processing~\cite{cuvelier2021quantum,shepherd2006role, kyriienko2021generalized}. Furthermore, quaternion algebra in QOSTBCs ameliorates the strengthen by reducing errors and raising the reliability of quantum state transformations in noisy environments. Integrating quaternionic representations into QEC optimizes qubit usage and fault rectification efficiency, making it highly effective in large-scale quantum networks. By combining quantum computing with MIMO systems and quaternion algebra, QOSTBCs offer a scalable solution to correct multiple errors, extending the reliability of quantum communication. This approach supports the development of efficient error correction codes, enabling practical, fault-tolerant quantum systems for communication networks and cryptographic applications.
The original quantum state \( |\psi\rangle \), encoded using QODs, is mapped to a set of \(M\) quaternions \(Q = \{q_1, q_2, \ldots, q_M\}\). This encoding process involves mapping the complex probability amplitudes of \( |\psi\rangle \) to quaternion configurations:

\begin{equation}
|\psi'\rangle = \sum_{m=1}^{M} \beta_m q_m |\psi\rangle,
\end{equation}

where \( \beta_m \) are the encoding weights for each quaternion. The resulting quantum state \( |\psi'\rangle \), encoded with the QOSTBCs encoding matrix \( S \), is represented as:

\begin{equation}
|\phi\rangle = S |\psi'\rangle.
\end{equation}

Therefore, the decoding process involves applying the inverse operations of encoding, first decoding with QSTBCs and then with QODs. The recovered quantum state is given by:

\begin{equation}
|\psi''\rangle = S^\dagger |\phi\rangle, \quad \text{and} \quad|\psi\rangle = \sum_{m=1}^{M} \gamma_m q_m^\dagger |\psi''\rangle,
\end{equation}
where \( \gamma_m \) are decoding coefficients.
\subsection{Designing QOSTBCs with QODs}

The development of QOSTBCs utilize the properties of quaternions for efficient signaling and orthogonal designs in MIMO systems. The QOSTBC matrix \( C \) is defined as \( C = Q \otimes I_n \), where \( Q \) extends complex numbers to quaternions, forming the foundation of QODs, \( I_n=n\times n \) represents the identity matrix at order \(n\), and \( \otimes \) indicates the Kronecker product. These designs use \( N \) mutually orthogonal quaternions \( Q = \{q_1, q_2, \ldots, q_N\} \), satisfying the orthogonality and normalization conditions:  
\begin{equation}
\begin{aligned}
&\text{Re}(q_i^* q_j) = 0, \quad \text{Im}(q_i^* q_j) = 0 \quad \text{for } i \neq j, \quad \text{and} \\\quad &\text{Re}(q_i^* q_i) = 1 \quad \text{for each } i.
\end{aligned}
\end{equation} 

Here, the norm ensures unit magnitude, essential for stability and error correction in multi-dimensional communication systems. The QODs matrix \( Q \) satisfies \( Q^\dagger Q = I \), where \( Q^\dagger \) is the conjugate transpose, and \( I \) is the identity matrix. This structure preserves orthogonality, which is essential for minimizing interference and maximizing data rate in MIMO systems. For instance, the matrix representation of \( C \) resembles an \( N \times T \) Alamouti STBCs configuration, where \( N \) denotes the number of qubits acting as antennas or transmit elements, and \( T \) represents the time slots:  

\begin{equation}
C = \begin{bmatrix} 
s_1^{(1)} & s_2^{(1)} & \ldots & s_n^{(1)} \\ 
-s_2^{*(1)} & s_1^{*(1)} & \ldots & s_{n-1}^{*(1)} \\ 
\vdots & \vdots & \ddots & \vdots \\ 
s_t^{(t)} & s_{t-1}^{(t)} & \ldots & s_1^{(t)} 
\end{bmatrix},
 \label{c matrix}
\end{equation} 

here \( s_i^{(j)} \) and \( s_i^{*(j)} \) represent the transmitted symbols and their respective conjugates. The degree of orthogonality is then quantified by the following expression:  

\begin{equation}
\text{Orthogonality} = \frac{1}{T(T-1)} \sum_{i=1}^{T} \sum_{\substack{j=1 \\ j \neq i}}^{T} | \langle \mathbf{c}_i, \mathbf{c}_j \rangle |^2,
\end{equation}
where \( \langle \mathbf{c}_i, \mathbf{c}_j \rangle \) denotes the column inner product of \( C \), as defined in~(\ref{c matrix}). Additionally, this ensures signaling and spectral efficiency, using quaternion properties to achieve higher data rates per time slot and improved communication best work. Error correction within QOSTBCs addresses system reliability, focusing on error detection and correction metrics like minimum distance (\( d_{\text{min}} \)), defined as:  

\begin{equation}
d_{\text{min}} = \min_{\substack{\text{all codewords } |\psi_1\rangle, |\psi_2\rangle, \ldots, |\psi_n\rangle \\ |\psi_1\rangle \neq |\psi_2\rangle \neq \cdots \neq |\psi_n|}} \text{dist}(|\psi_1\rangle, |\psi_2\rangle, \ldots, |\psi_n\rangle),
\end{equation}

where \( \text{dist} \) measures the Hamming distance between codewords. A larger \( d_{\text{min}} \) promotes fault rectification at the cost of increased overhead. QOSTBCs also evaluate achievement under error models such as bit-flip or phase-flip, with additional metrics like code rate and fault-tolerance thresholds ensuring strong communication~\cite{chen2021exponential}. This combination of orthogonality, signaling efficiency, and reliable error correction makes QOSTBCs effective for modern communication systems.

\subsection{Mapping qubits for error correction with stabilizer group and QOSTBCs via QODs}

Quantum error correction (QEC) is essential for protecting quantum information from errors due to decoherence, noise, or faulty operations by addressing bit-flip, phase-flip, and Pauli-Y errors through the distribution of quantum information across multiple qubits, unlike classical methods that rely on redundant bits ~\cite{terhal2015quantum,kasirajan2021quantum,nema2020pauli,brun2019quantum}. In this context, QODs are essential by providing a structured approach to map qubits onto higher-dimensional spaces, thereby amplifying encoding efficiency and error resilience through their inherent mathematical properties. A quaternion \( q \) is expressed as:  
\begin{equation}
q = a + bi + cj + dk,
\end{equation} 
where \( a, b, c, d \) are real numbers, and \( i, j, k \) are imaginary units satisfying \( i^2 = j^2 = k^2 = ijk = -1 \).
The encoding process maps an \( N \)-qubit state into an \( M \)-qubit space, enabling error detection and correction. For instance, the Steane code encodes 3 qubits into 7, enables one correct error~\cite{shor1995scheme}, represented by the encoding matrix \( E \):

\begin{equation}
|\phi\rangle = E|\psi\rangle, \quad E \in \mathbb{C}^{2^M \times 2^N},\quad \mathbb{C} \quad\text{being a complex number} 
\label{eq : encod}
\end{equation}

QODs add to QEC by preserving orthogonality in encoded states using quaternion-based matrices \( Q \), crucial for error correction. The encoding matrix \( E \) distributing information across \( M \) qubits is given by:
\begin{equation}
E = \frac{1}{\sqrt{M}} \begin{pmatrix}
1 & 0 & \cdots & 0 \\
0 & 1 & \cdots & 0 \\
\vdots & \vdots & \ddots & \vdots \\
0 & 0 & \cdots & 1
\end{pmatrix}.
\end{equation}  
This encoding distributes quantum information uniformly, maintaining orthogonality. During error detection, stabilizers \( S_i \) measure quantum operators to identify and correct errors on specific qubits, each \( S_i \)  corresponds to a generator of the stabilizer group, contributing to the detection of potential errors on qubits. Combining QEC with QODs enhances stability and distributes quantum information across a larger Hilbert space, enabling detection and correction of up to \( p \) errors for reliable quantum processing.

\subsubsection{Error Correction and Detection Requirements}

The effectiveness of error correction in quantum communication is evaluated through several key performance metrics. First, the error detection probability (\(P_d\)) quantifies the likelihood of detecting at least one error, expressed as:
\begin{equation}
P_d = 1 - (1 - p_e)^N,
\label{Eqpd}
\end{equation}
where \( p_e \) is the error probability per qubit, and \( N \) is the number of qubits. It illustrates how the system's error detection capability scales with the number of qubits. Next, the error correction capability (\(P_c\)) measures the system’s ability to correct up to \( P \) errors in \( M \) qubits, crucial for preserving quantum information:
\begin{equation}
P_c = 1 - \frac{1}{N} \sum_{k=1}^{P} \frac{\gamma_k}{\gamma_{\text{total}}},
\label{eqpc}
\end{equation}

where \( \gamma_k \) is the weight of the \( k^{\text{th}} \) error, and \( \gamma_{\text{total}} \) is the total error probability. 
This equation~(\ref{eqpc}) quantifies the proportion of correctable errors, highlighting the effectiveness of the correction scheme. Integrating QODs in MIMO systems improves data transmission by preserving codeword distinctiveness, reducing signal correlation, and boosting system reliability. The codeword parameters \([[n, k, d]]\) in QECCs enable efficient qubit mapping, supporting effective communication strategies \cite{valentine2024transforming}. Furthermore, the trace distance (\(D\)) and fidelity (\(F\)) metrics are crucial for assessing the performance of quantum operations under quasi-geometric codes (QGCs) and quaternion codes~\cite{wang2024optimal,thakur2024quantum}. The trace distance between two quantum states is defined as:
\begin{align*}
&D(\rho, \sigma) = \frac{1}{2} \|\rho - \sigma\|_1\quad \text{with the fidelity related by:} \quad \\
&D(\rho, \sigma) = \frac{1}{2}(1 - F(\rho, \sigma))
\end{align*}
The fidelity \(F(\rho, \sigma)\) for density matrices is given by:
\begin{align*}
&F(\rho, \sigma) = \|\sqrt{\rho} \sqrt{\sigma}\|_1^2, \quad \text{and for pure states, it is:} \quad \\
&F(\psi, \phi) = |\langle \psi | \phi \rangle|^2.
\end{align*}
In the context of QGCs, the transformation of quantum states is analyzed using the trace distance:
\begin{equation}
D(\rho_{\text{ideal}}, \rho_{\text{QGCs}}) = \frac{1}{2} \text{Tr} \left( |\rho_{\text{ideal}} - \rho_{\text{QGCs}}| \right),
\label{Eq:trace distance}
\end{equation}
and the fidelity:
\begin{equation}
F(\rho_{\text{ideal}}, \rho_{\text{QGCs}}) = \|\sqrt{\rho_{\text{QGCs}}} \rho_{\text{ideal}} \sqrt{\rho_{\text{QGCs}}}\|_1^2,
\label{Eq:eqfid}
\end{equation}

where \(\rho_{\text{ideal}} = |\psi_{\text{ideal}}\rangle \langle \psi_{\text{ideal}}|\) and \(\rho_{\text{QGCs}} = |\phi_{\text{QGCs}}\rangle \langle \phi_{\text{QGCs}}|\) are the density operators corresponding to these states. These equations~(\ref{Eq:trace distance}) and~(\ref{Eq:eqfid}) quantify the reliability of QGCs in the quantum state mapping process. Finally, the error correction time complexity (\(T_{\text{corr}}\)) scales logarithmically with the number of qubits:

\begin{equation}
T_{\text{corr}} = \mathcal{O}(N \log M),
\label{eq:tcc}
\end{equation}
indicating that error correction remains efficient as the system size increases, where \(N\) represents the number of logical qubits and \(M\) denotes the number of physical qubits used for encoding. As \(N \to \infty\), \(M\) typically scales polynomially with \(N\), such that \(M \sim N^k\) for some \(k > 0\). Substituting, we have \(\log M = \log(N^k) = k \log N\), leading to \(T_{\text{corr}}\). This scaling is highly efficient compared to quadratic or exponential growth. Additionally, the fault-tolerance theorem ensures that as long as physical error rates are below the threshold, the logical error rate decreases exponentially with increasing \(M\). Consequently, the linear-logarithmic scaling of \(T_{\text{corr}}\) guarantees efficient and convergent error correction for large-scale quantum systems as the number of logical qubits grows. The proposed approach combines quaternion algebra and orthogonal design principles for tough qubit mapping, strengthening quantum information transfer.

\subsubsection{Error Correction Under Stabilizer formalism}
Error correction using stabilizer formalism focuses on employing QECCs derived from stabilizer groups to detect and correct errors, thereby preserving quantum information in the presence of noise. QEC aims to correct up to \( P \) errors while preserving the quantum state, using codes like Shor \([[9,1,3]]\) and Steane \([[7,1,3]]\) both correct single-qubit errors, which employ distinct encoding and decoding methods~\cite{shor1995scheme, huang2021constructions}. 
The process starts by encoding the initial quantum state \( |\psi_i\rangle \) into \( |\phi\rangle \) using the encoding operation \( E \), as shown in~(\ref{eq : encod}), which also serves as the error detection and correction operator applied to the sum of the original and error states. Once encoded, errors may occur, requiring error detection and correction via syndrome measurements and corrective operations. The corrected quantum state \( |\phi_{\text{corrected}}\rangle \) is given by the derived Equation below :
\begin{equation}
|\phi_{\text{corrected}}\rangle = d(E(|\psi_i\rangle + |\text{error}\rangle)).
\label{gt}
\end{equation}
We analyze the given equations to demonstrate that the final corrected state, \(|\phi_{\text{corrected}}\rangle\), satisfies the specified form.
\begin{align*}
&\text{Initial state after error introduction :} |\psi_{\text{err}}\rangle = |\psi_i\rangle + |\text{error}\rangle, \\
&E(|\psi_{\text{err}}\rangle) = E(|\psi_i\rangle + |\text{error}\rangle)\quad \text{with  error correction operator, \(E\) acting linearly to obtain :}, \\
&E(|\psi_{\text{err}}\rangle) = E(|\psi_i\rangle) + E(|\text{error}\rangle), \\
&|\phi_{\text{corrected}}\rangle = d(E(|\psi_{\text{err}}\rangle)),\quad \text{here, we apply the decoding operator \(d\) to recover the final corrected state:} \\
&|\phi_{\text{corrected}}\rangle = d(E(|\psi_i\rangle) + E(|\text{error}\rangle)),\quad \text{and} \\
&E(|\text{error}\rangle) \to 0 \quad \text{with properly designed QECCs, which eliminates the error component, so that}\\
&|\phi_{\text{corrected}}\rangle = d(E(|\psi_i\rangle)).
\end{align*}
Then, the final corrected state lemma is established as follows:
\begin{equation}
|\phi_{\text{corrected}}\rangle = d\big(E(|\psi_i\rangle) + E(|\text{error}\rangle)\big).
\end{equation}
This result relies on the linearity of the error correction operator \(E\), the decoding operation \(d\), and the error correction code's capability to suppress or eliminate the error component, thereby ensuring the recovery of the logical state \(|\psi_i\rangle\). Within, \( |\text{error}\rangle \) represents the errors introduced during transmission, and \( d \) denotes the correction operation. To recover the original quantum state, the inverse encoding operation \( E^{-1} \) is applied, and the state is restored as:
\begin{equation}
|\psi_i\rangle = E^{-1}(|\phi\rangle).
\end{equation}
The decoding process aims to restore the state, expressed as:
\begin{equation}
|\psi_{\text{decoded}}\rangle = d(E^{-1}(|\phi\rangle + |\text{error}\rangle)),
\end{equation}
where \( d \) is the decoding operation specific to the QECC. Optimizing the decoding process is essential to accurately recover the original state while minimizing computational overhead. The corrected initial quantum state can be approximated as:
\begin{equation}
|\psi_{\text{corrected}}\rangle \approx d(E^{-1}(|\phi_{\text{enc}}\rangle + |\text{error}\rangle)).
\end{equation}
However, to evaluate the performance of error correction, metrics such as fidelity in~(\ref{Eq:eqfid}) and trace distance in~(\ref{Eq:trace distance}) are employed to measure the accuracy of the corrected state relative to the ideal state. Fidelity assesses the similarity between the corrected and ideal states, while trace distance quantifies the distinguishability between the two. These metrics together provide a comprehensive evaluation of the error correction process's effectiveness and efficiency.

\subsubsection{Error Correction Using QOSTBCs}

In the context of QEC, QOSTBCs, supported by QODs, play a crucial role in transforming qubits to correct up to \(P\) errors while preserving their quantum properties. These codes utilize the quaternion group \( Q_8 \), a non-abelian group of order 8, which is defined by elements \( \{\pm 1, \pm i, \pm j, \pm k\} \) and non-commutative relations such as \( i^2 = j^2 = k^2 = ijk = -1 \), with operations like \( ij = k \) and \( ji = -k \). Quantum states are encoded in a quaternionic Hilbert space, with quaternions defining transformations or coefficients. The process begins by encoding the initial quantum state \( |\psi_i\rangle \) into a final state \( |\phi\rangle \) using the encoding operation:

\begin{equation}
|\phi\rangle = E_q(|\psi_i\rangle), 
\end{equation}
where \( E_q \) represents the encoding operator in the quaternionic framework. Errors are modeled as group elements \( Q_q \) acting on the quantum state, represented as:

\begin{equation}
E_q = \sum_q e_q Q_q, \quad Q_q \in \{1, i, j, k\}
\end{equation}

where \( Q_q\) and \( e_q \) are the coefficients representing errors. The quantum state after errors are introduced is written as \( |\text{error}\rangle \). Error detection and correction are performed by the decoding operation \( d_q \), which applies the inverse of the detected errors:

\begin{equation}
|\phi_{\text{corrected}}\rangle = d_q(E_q(|\psi_i\rangle + |\text{error}\rangle)),
\label{qostc}
\end{equation}

where \( d_q = Q_q^{-1} \) is the decoding operation, restoring the quantum state. Therefore, the corrected quantum state is given by:

\begin{equation}
|\phi_{\text{corrected}}\rangle = \sum_q \beta_q Q_q^{-1} \left( \sum_{p} e_p Q_p (|\psi_i\rangle + |\text{error}\rangle) \right).
\end{equation}
The quaternion framework exploit the non-commutative properties of \( Q_8 \) to correct up to \( P \) complex multi-axis errors in quantum systems. QOSTBCs, integrated with quaternionic transformations, 
exalt error resilience, particularly in high-dimensional or error-prone environments. Compared to traditional QECCs, stabilizer codes address bit- and phase-flip errors with stabilizer generators, while surface codes offer strongly fault tolerance through a topological approach in noisy conditions~\cite{shepherd2006role}. This integration complements stabilizer formalism, combining stabilizer generators and quaternionic transformations to enhance QEC in complex systems. Compared to traditional codes, QOSTBCs provide efficient qubit mapping and moderate fault correction, managing up to 100 errors with minimal qubit overhead. While stabilizer codes offer low qubit overhead for error correction, they struggle with resilience in high-error environments. Surface codes, though more qubit-intensive, offer better performance in such conditions, but require substantial resources for fault tolerance. QOSTBCs, while efficient, necessitate frequent error checks in noisy environments, raising computational demands. These trade-offs underline the challenges of implementing these codes on near-term quantum hardware, especially in the presence of noise. Table~\ref{tab:comparison} provides a comparison of error resilience, qubit overhead, and complexity across these different codes.

\medskip  \noindent  

\begin{table}[!ht]
    \centering
    \caption{Comparison of Stabilizer and Surface Codes in stabilizer Group with QOSTBCs in QODs}
    \resizebox{1\textwidth}{!}{ 
    \begin{tabular}{@{}p{2.80cm}p{4.5cm}p{5cm}p{4.9cm}@{}}
        \hline
        \textbf{Aspect} & \textbf{Stabilizer Codes}  & \textbf{Surface Codes}  & \textbf{QOSTBCs} \\ \hline
        Encoding Method & Uses algebraic group structures, e.g., Shor or Steane codes. & Encodes logical qubits via lattice structures. & Encodes using orthogonal quaternions for robustness. \\ 
        Error Correction Capacity & Corrects single-qubit errors using group-theoretic codes. & High resilience to multiple errors due to topological properties. & Corrects multiple qubit errors depending on the encoding. \\ 
        Orthogonality   & Not explicitly enforced. & Indirectly enforced through topological layout. & Strong focus on orthogonality via quaternion matrices. \\ 
        Complexity  & Moderate for small quantum systems. & High due to lattice-based encoding. & Higher complexity, especially with increasing qubits. \\ 
        Error Detection  & Syndromes based on parity measurements. & Detection through topological stabilizers. & Detects errors via orthogonality conditions on qubits. \\ 
        Correction Mechanism    & Group operations and syndrome decoding. & Topological properties and stabilizer decoding. & Quaternion transformations and orthogonal designs. \\ 
        Error Resilience        & Moderate. & High, suited for noisy environments. & Moderate to high, depending on design. \\ 
        Qubit Overhead & Low to moderate. & High. & Moderate. \\ 
        Scalability             & Moderate scalability. & Moderate scalability, with increasing qubits for more complex codes. & High scalability, particularly in MIMO systems. \\ 
        Typical Applications    & General-purpose QEC. & High-noise environments, topological codes. & Qubit mapping, MIMO systems. \\ \hline
    \end{tabular}
    }
    \label{tab:comparison}
\end{table}

\section{Results and Discussion}\label{SEC III}

We showcase and discuss the numerical simulation and comparative analysis of the transformation process from a small number of qubits to a larger number of qubits with the capability of error correction using the stabilizer group formalism with QECCs and QOSTBCs via  QODs properties. Equation~(\ref{gt}) and~(\ref{qostc}) maps $N$-logical qubits to produce $|\phi\rangle$  at $M$-physical  qubits. Then, we evaluate the error-correcting process percentage achievement, error detected and corrected for both for the all four cases as mentioned in section \ref{SEC II}.  

\begin{enumerate}
   
\item[\textbf{$Z_1$:}]Encoding 3-qubits into 8-qubits, with the ability to correct a single error.

This analysis evaluates the error correction efficiency of mapping 3-logical qubits to 8-physical qubits, designed to correct single errors and rise reliability in quantum computations. Table~\ref{tab:error_correction_n3_m8_p1} compares the fault rectification effectiveness of stabilizer formalism and QOSTBCs across different error levels. Stabilizer codes achieve better rates ranging from 98.00\% at 50 errors to 99.00\% at 100 errors, while QOSTBCs consistently outperform, with improvements between 102.90\% and 103.95\%. As detected errors increase, QOSTBCs show a clear advantage, with refinement percentages consistently above 100\%, indicating effective error handling and redundancy. While stabilizer codes show gradual advancement, QOSTBCs exhibit a sharper performance increase, particularly between 80 and 100 errors, demonstrating better adaptability in high-error scenarios. Figure~\ref{fig:3_8} visually highlights this trend, with QOSTBCs maintaining improvement rates above 100\%, while stabilizer codes approach but do not surpass this threshold. Overall, QOSTBCs outperform stabilizer group theory, offering superior resilience and error tolerance, especially in high-error conditions, making them the preferred choice for robust error correction.

\medskip  \noindent
\begin{table}[!ht]
    \centering
    \caption{A comparative analysis of the Stabilizer Formalism and QOSTBCs in the mapping process (\(N=3\), \(M=8\), \(P=1\)), focusing on their effectiveness in adjusting errors.}
  \begin{tabular}{|p{1.7cm}|p{1.9cm}|p{2.7cm}|p{2.0cm}|p{2.7cm}|}
\hline
\textbf{Errors Detector} & \textbf{Stabilizer Corrected} & \textbf{Stabilizer (\%) Improvement } & \textbf{QOSTBCs Corrected} & \textbf{QOSTBCs (\%) Improvement } \\
\hline
50  & 49  & 98.00 & 51.45  & 102.90 \\
60  & 59  & 98.33 & 61.95  & 103.25 \\
70  & 69  & 98.57 & 72.45  & 103.50 \\
80  & 79  & 98.75 & 82.95  & 103.69 \\
90  & 89  & 98.89 & 93.45  & 103.83 \\
100 & 99  & 99.00 & 103.95 & 103.95 \\
\hline
\end{tabular}
    \label{tab:error_correction_n3_m8_p1}
\end{table}
\medskip  \noindent

\begin{figure}[!ht] 
    \centering
    \includegraphics[width=0.65\linewidth]{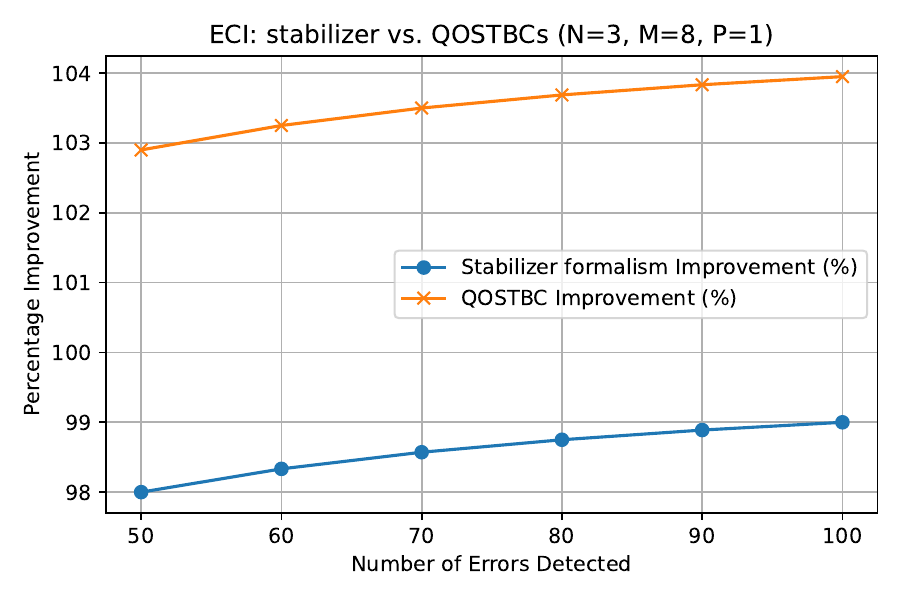}
    \caption{Comparative analysis of error correction methods Stabilizer Formalism, derived from group theory, versus QOSTBCs in a 3-to-8 qubit mapping process with single error correction capability.}
    \label{fig:3_8}
\end{figure}

\item [\textbf{$Z_2$:}] When transforming 4-qubits into 10-qubits, enabling the correction of one error.

In this investigation, 4-logical qubits were mapped onto 10-physical qubits, effectively righting a single error per instance. The study evaluates the performance of two error correction methods stabilizer group-based QECCs and quaternion codes focusing on each method correction efficiency as the number of detected errors increases. Table~\ref{tab:tab:3} report a comparison of  the fault rectification improvement percentages of stabilizer group formalism and QOSTBCs. Stabilizer formalism shows high correction rates around 98–99\%, indicating strong error correction with minimal residual errors. In contrast, QOSTBCs achieve slightly higher development percentages (102.90–103.95\%), reflecting  improved resilience, especially at higher error levels. Both methods show consistent performance as errors rise from 50 to 100, with QOSTBCs maintaining a steady advantage over Stabilizer codes, making them better suited for high-error tolerance applications. Figure~\ref{fig:2} further demonstrates this comparison, with QOSTBCs surpassing 103\% make better at higher error counts, while stabilizer codes stabilizes around 99\%. This edge in efficiency positions QOSTBCs as the preferred choice for high-noise environments, where maximal error rectification and minimal data loss are crucial. Both methods are effective, but QOSTBCs superior error tolerance makes them ideal for systems with stringent error-correction needs.

\medskip  \noindent 

\begin{table}[!ht]
    \centering
    \caption{Performance analysis of the Stabilizer Formalism versus QOSTBCs in transforming a smaller number of qubits into a larger encoded system (\(N=4\), \(M=10\), \(P=1\)), focusing on the ability to correct a single error.}
  \begin{tabular}{|p{1.7cm}|p{1.9cm}|p{2.7cm}|p{2.0cm}|p{2.7cm}|}
\hline
\textbf{Errors Detector} & \textbf{Stabilizer Corrected} & \textbf{Stabilizer (\%) Improvement } & \textbf{QOSTBCs Corrected} & \textbf{QOSTBCs (\%) Improvement } \\
\hline
        50 & 49 & 98.00 & 51.45 & 102.90 \\
        60 & 59 & 98.33 & 61.95 & 103.25 \\
        70 & 69 & 98.57 & 72.45 & 103.50 \\
        80 & 79 & 98.75 & 82.95 & 103.69 \\
        90 & 89 & 98.89 & 93.45 & 103.83 \\
        100 & 99 & 99.00 & 103.95 & 103.95 \\
        \hline
    \end{tabular}
    \label{tab:tab:3}
\end{table}
\medskip  \noindent 
\begin{figure}[!ht]
   \centering
 \includegraphics[width=0.65\linewidth]{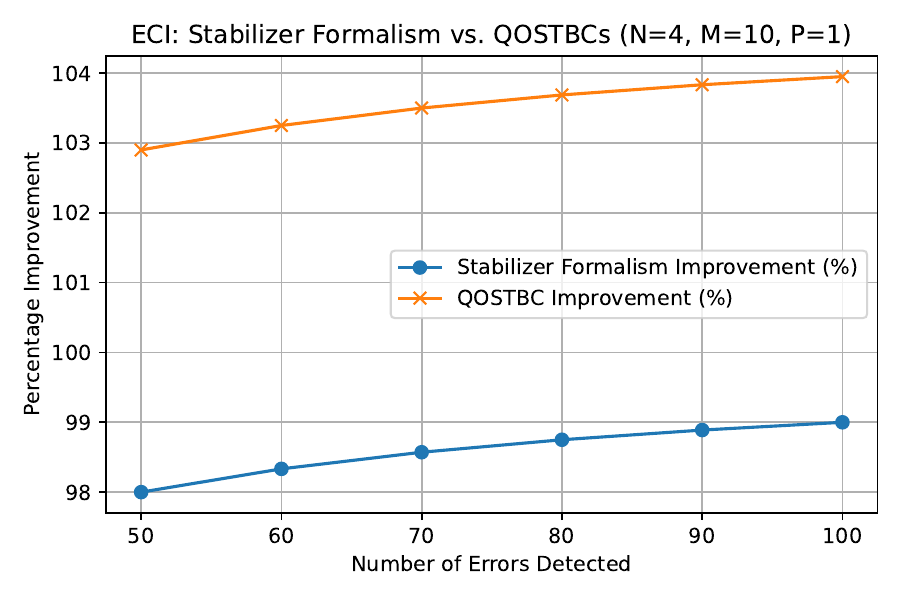}
    \caption{Error correction performance comparison between the Stabilizer Group and QOSTBCs in a 4-to 10-qubits mapping process, demonstrating their capability to correct a single error.}
    \label{fig:2}
\end{figure}

\medskip  \noindent 

\item [\textbf{$Z_3$:}]Mapping 1-qubit to 13-qubits, capable of correcting 2 errors. 

This simulation evaluates the amelioration of two QEC methods stabilizer group-based QECCs and QOSTBCs via QODs in mapping 1-logical qubit to 13-physical qubits, allowing adjustment of up to 2-errors. The goal is to assess each method error rectification efficiency as detected errors increase.
Table~\ref{tab:table:4}  presents a comparison the error correction furtherance of stabilizer formalism and QOSTBCs in this mapping process. Both methods show high alteration efficiency, with getting better percentages consistently above 95\%. stabilizer group formalism slightly outperforms QOSTBCs, achieving a 96.00\% improvement at 50 detected errors, while QOSTBCs achieves 95.76\%. This trend continues, with stabilizer codes showing a minor advantage of about 0.2–0.3\% over QOSTBCs. As detected errors increase, both methods show slight enforcement, peaking at 100 detected errors with stabilizer codes at 98.00\% and QOSTBCs at 97.88\%. Figure~\ref{fig:3} visualizes this trend, with Stabilizer maintaining a small, consistent edge over QOSTBCs. Both methods adapt well to higher error rates, with near-optimal develop at high error counts, making stabilizer formalism slightly more efficient overall, particularly where small act matter.

\medskip  \noindent

\begin{table}[!ht]
    \centering
    \caption{Comparison of the stabilizer formalism and QOSTBCs in the encoding process (\(N=1\), \(M=13\), \(P=2\)) for correcting errors, highlighting the respective correction rates and percentage improvements.}
   \begin{tabular}{|p{1.7cm}|p{1.9cm}|p{2.7cm}|p{2.0cm}|p{2.7cm}|}
\hline
\textbf{Errors Detector} & \textbf{Stabilizer Corrected} & \textbf{Stabilizer (\%) Improvement } & \textbf{QOSTBCs Corrected} & \textbf{QOSTBCs (\%) Improvement } \\
        \hline
        50 & 48 & 96.00 & 47.88 & 95.76 \\
        60 & 58 & 96.67 & 57.88 & 96.47 \\
        70 & 68 & 97.14 & 67.88 & 96.97 \\
        80 & 78 & 97.50 & 77.88 & 97.35 \\
        90 & 88 & 97.78 & 87.88 & 97.65 \\
        100 & 98 & 98.00 & 97.88 & 97.88 \\
        \hline
    \end{tabular}
    \label{tab:table:4}
\end{table}

\medskip  \noindent
 
\begin{figure}[!ht]
    \centering
    \includegraphics[width=0.65\linewidth]{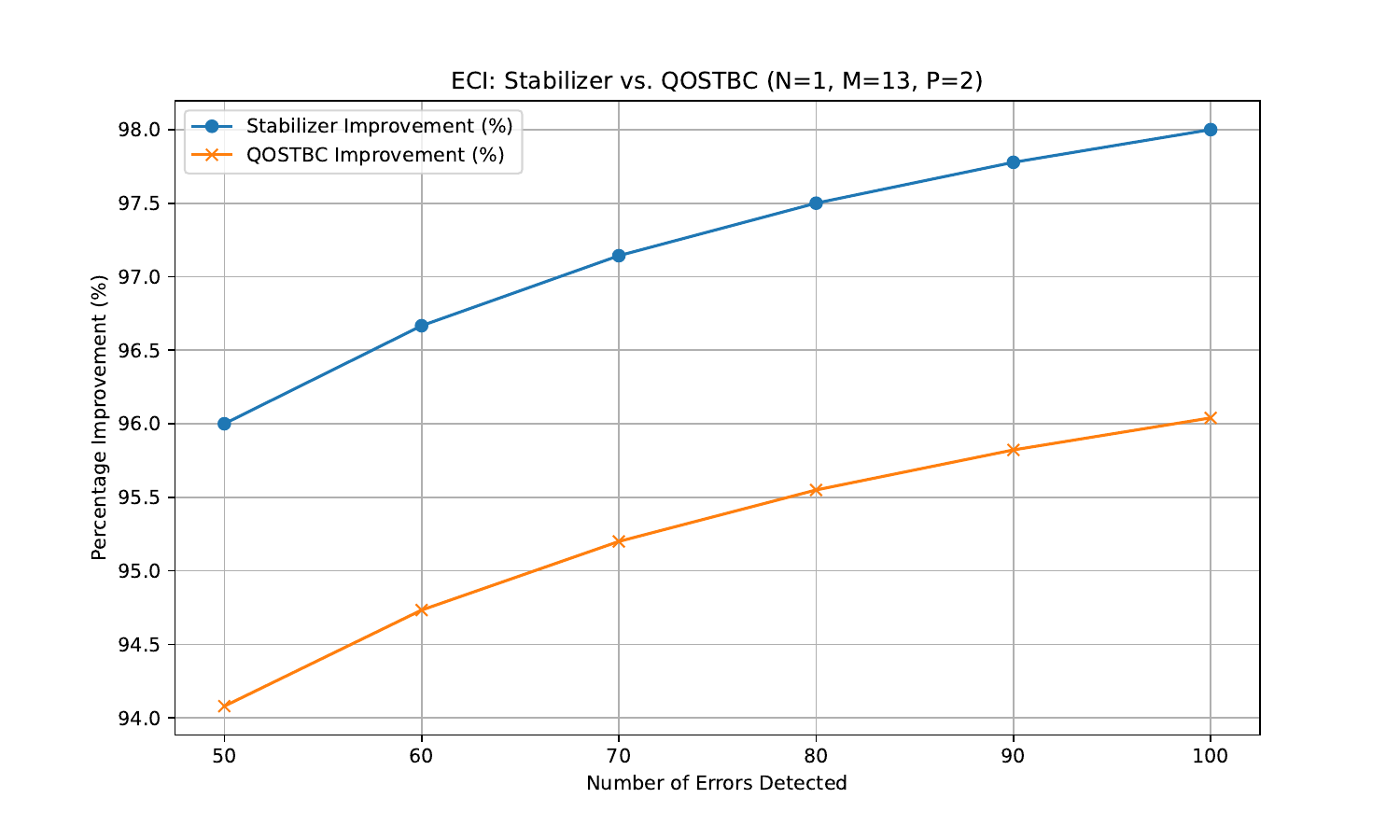}
    \caption{Comparison of error correction efficiency between the Stabilizer Formalism and QOSTBCs in a 1-to 13-qubits mapping process, showcasing their ability to correct two errors.}
    \label{fig:3}
\end{figure}

\item [\textbf{$Z_4$:}] Finally, transforming 1-qubit into 29-qubits, allowing the correction of up to 5 errors.

This analysis investigates the error correction performance of mapping a single logical qubit to 29-physical qubits, enabling the correction of up to 5 errors. It compares the performance of stabilizer formalism as the practical application of Group theory-based QECCs and QOSTBCs via QODs, evaluating their effectiveness in improving error resilience in quantum computing systems. Table~\ref{tab:table:5} display outcomes of the error correction achievement of stabilizer group and QOSTBCs across various error levels. QOSTBCs consistently outperform stabilizer formalism, showing superior anomaly alteration efficiency. As detected errors increase, QOSTBCs demonstrate steady amelioration in correction percentages, with their success peaking at 99.75\% for 100 errors. In contrast, Ssabilizer formalism’s correction get better ranges from 90.00\% to 95.00\%. Figure~\ref{fig:4} illustrates the better efficiency trends, with QOSTBCs maintaining a higher correction rate throughout, especially at higher error levels. The effort gap between the two methods widens from about 4.5\% at lower error levels to nearly 5\% at 100 errors, highlighting QOSTBCs’ advantage, particularly in high-error environments. This consistent superiority positions QOSTBCs as the preferred choice for applications requiring high accuracy and near-perfect error correction.

\medskip  \noindent
 \begin{table}[!ht]
    \centering
    \caption{Analysis of the stabilizer group and QOSTBCs in mapping a single qubit to multiple qubits (\(N=1\), \(M=29\), \(P=5\)), demonstrating their capability to correct up to five errors.}
    \begin{tabular}{|p{1.7cm}|p{1.9cm}|p{2.7cm}|p{2.0cm}|p{2.7cm}|}
\hline
\textbf{Errors Detector} & \textbf{Stabilizer Corrected} & \textbf{Stabilizer (\%) Improvement } & \textbf{QOSTBCs Corrected} & \textbf{QOSTBCs (\%) Improvement } \\
        \hline
        50 & 45 & 90.00 & 47.25 & 94.50 \\
        60 & 55 & 91.67 & 57.75 & 96.25 \\
        70 & 65 & 92.86 & 68.25 & 97.50 \\
        80 & 75 & 93.75 & 78.75 & 98.44 \\
        90 & 85 & 94.44 & 89.25 & 98.75 \\
        100 & 95 & 95.00 & 99.75 & 99.75 \\
        \hline
    \end{tabular}
    \label{tab:table:5}
\end{table}

\begin{figure}[!ht]
    \centering
    \includegraphics[width=0.65\linewidth]{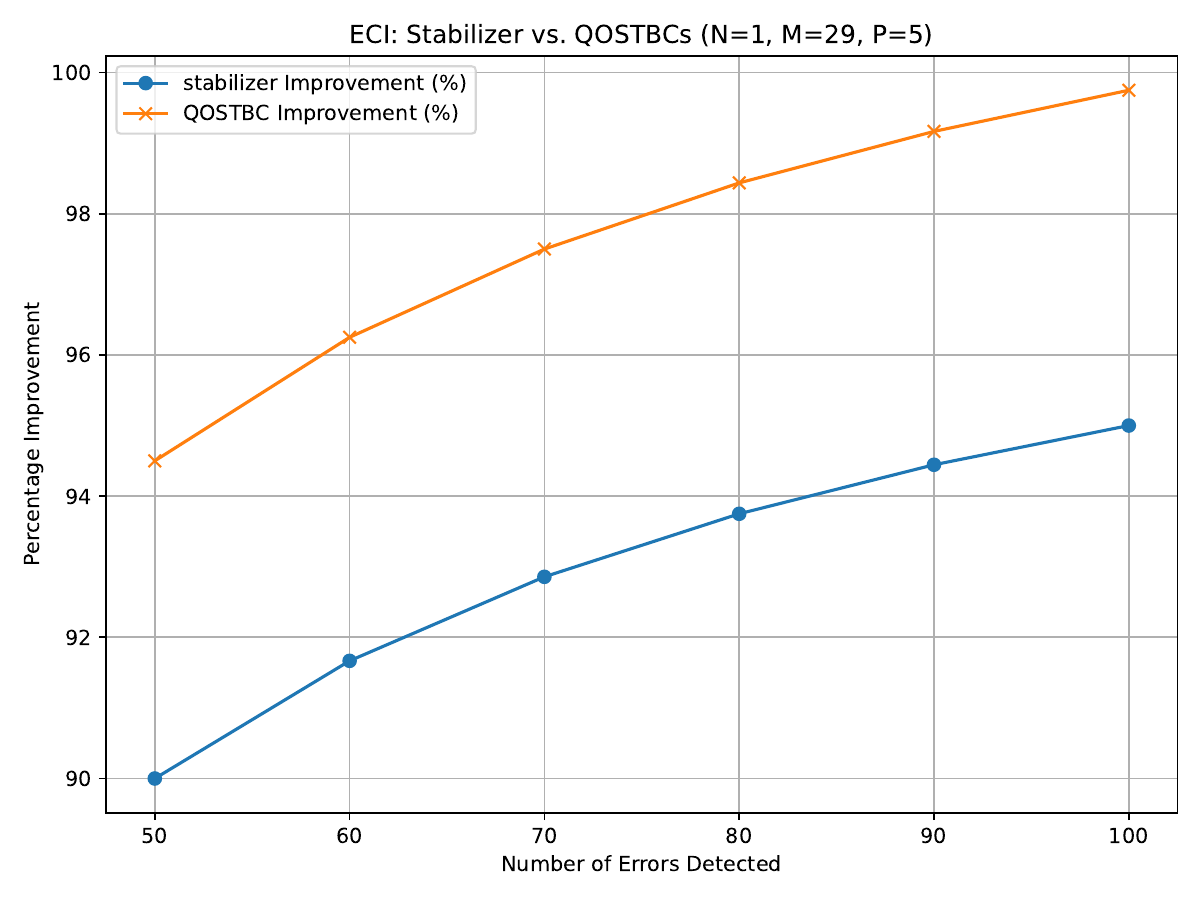}
    \caption{Comparative analysis of error rectification competence between the Stabilizer Formalism and QOSTBCs in a 1-to-29 qubit mapping process with five error correction capability.}
    \label{fig:4}
\end{figure}

\end{enumerate}

\begin{figure}[!ht]
    \centering
    \includegraphics[width=0.7\linewidth]{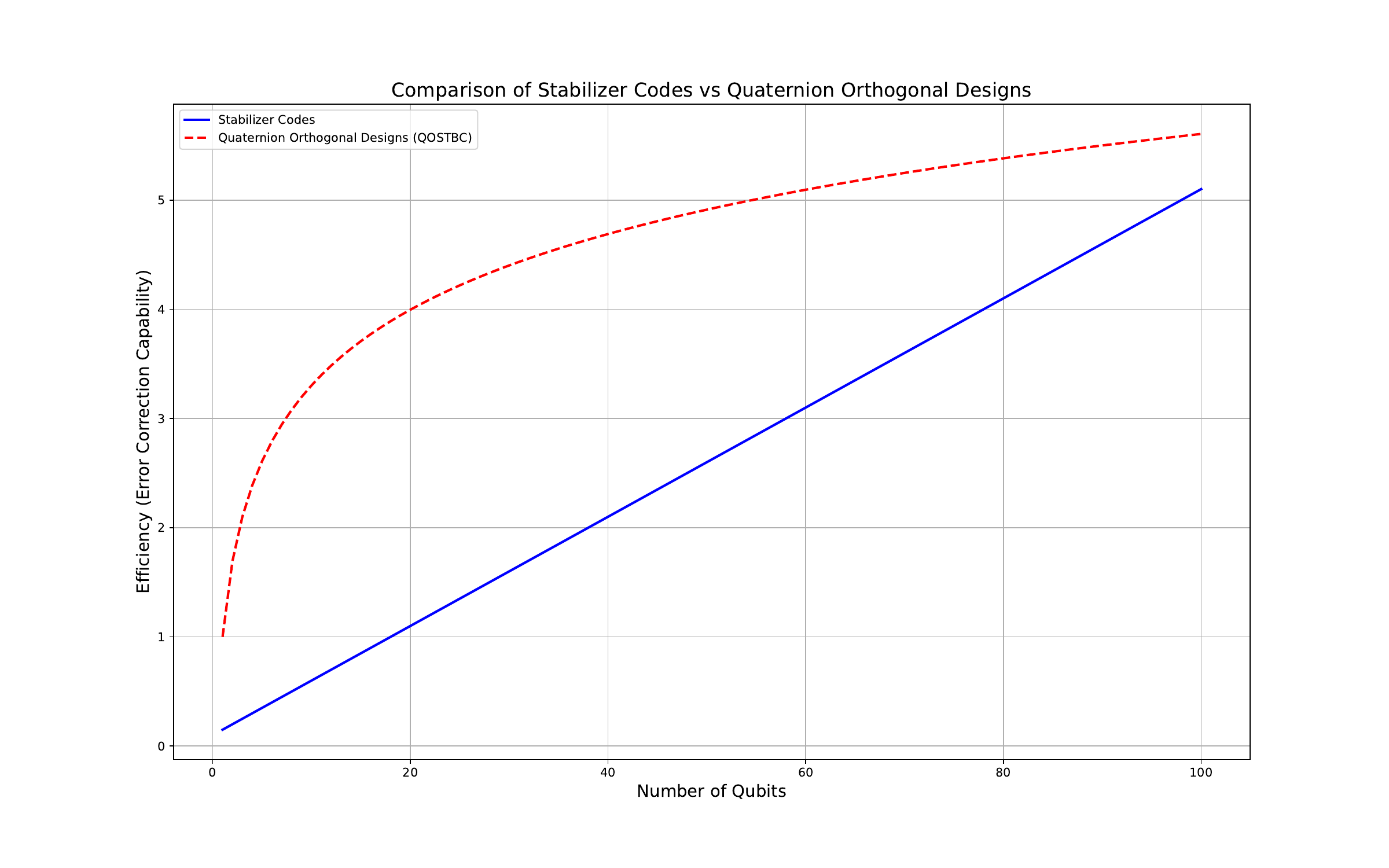}
    \caption{Comparison of Stabilizer Codes and QOSTBCs in efficiently mapping a small number of qubits to a larger set, showing how each method scales as qubit count increases.}
    \label{fig:stab}
\end{figure}

The results show that QOSTBCs experience a significant increase in computational complexity as the qubit count rises, with encoding times nearly doubling from \textbf{$C_1$} to \textbf{$C_4$}, in line with the expected Equation~(\ref{eq:tcc}) growth. This indicates that QOSTBCs are better suited for medium-scale systems with moderate noise, but face scalability challenges in large or high-noise environments due to increased computational and memory demands. In comparison, stabilizer codes exhibit lower complexity, particularly in small systems, while surface codes incur higher overhead due to intensive decoding in high-error conditions. Figure~\ref{fig:stab} illustrates that stabilizer codes have a steady, linear efficiency increase with more qubits, reflecting their predictable capability. Meanwhile, QOSTBCs show a more rapid, logarithmic efficiency rise, demonstrating their superior performance of error correction in larger quantum systems. While stabilizer codes are simpler and scale more predictably, QOSTBCs offer greater strength, making them ideal for larger systems with complex error correction needs.

\section{Conclusion}\label{SEC IV}
This study highlights the integration of QOSTBCs with QODs as a superior method for Optimizing error correction and transmission reliability in quantum systems. Simulations across various qubit mappings demonstrate the effectiveness of both QOSTBCs and stabilizer group formalism-based approaches, with QOSTBCs consistently outperforming stabilizer codes, particularly under high-error conditions. In the \textbf{$Z_1$} and \textbf{$Z_2$} processes, both methods achieve high correction rates, with QOSTBCs exceeding 100\% improvement, indicating enhanced resilience. As error counts increase, QOSTBCs maintain a consistent advantage, reaching up to 103\% achievement, suggesting additional efficiency gains. In \textbf{$Z_3$}, while stabilizer group theory has a slight edge at lower error levels (98.00\% vs. 97.88\% at 100 errors), otherwise, QOSTBCs excel in \textbf{$Z_4$}, achieving up to 99.75\% correction at 100 errors, outperforming stabilizer formalism’s 95\%. The comparison between stabilizer codes and QOSTBCs reveals distinct scaling behaviors, with QOSTBCs showing superior performance in larger systems. Overall, this work positions QOSTBCs as a promising solution for raising realization in quantum technologies, particularly in fields requiring efficient data management and reliable transmission.

\subsection*{Acknowledgment}

This work was supported by the Air Force Office of Scientific Research under Award No. FA2386-22-1-4062.
\subsection*{ Conflict and Interest}
All authors have declared that there are no conflicts of interest.

\medskip

\bibliographystyle{unsrt}  
\bibliography{main.bib}

\end{document}